\documentclass[12pt,preprint]{aastex}
\begin{document}
\title{High Frequency VLBI Imaging of the Jet Base of M87}
\shorttitle{VLBI IMAGING OF THE JET BASE OF M87}
\author{Chun Ly,\altaffilmark{1} R. Craig Walker,\altaffilmark{2} and William Junor\altaffilmark{3} }
\shortauthors{LY ET AL.}
\email{chun@astro.ucla.edu}

\altaffiltext{1}{Department of Astronomy, University of California, Los Angeles, P.O. Box 951547,
  Los Angeles, CA 90095-1547}
\altaffiltext{2}{National Radio Astronomy Observatory, P.O. Box 0, Socorro, NM 87801}
\altaffiltext{3}{ISR-2, MS-D436, Los Alamos National Laboratory, P.O. Box 1663, Los Alamos, NM 87545}
\begin{abstract}
  VLBA and Global VLBI observations of M87 at 43\,GHz, some new and some previously published, are
  used to study the structural evolution of the jet with a spatial resolution of under 100
  Schwarzschild radii. The images, taken between 1999 and 2004, have an angular resolution of
  0\farcs00043 $\times$ 0\farcs00021. An edge-brightened jet structure and an indication of a large
  opening angle at the jet base are seen in all five epochs. In addition, a probable
  counter-jet is seen in the latter three epochs.
  A 22\,GHz VLBA image also confirms many of the structures seen at the higher frequency, including
  the counter-jet. A comparison of the counter-jet flux density at 22 and 43\,GHz reveals that it is
  not free-free absorbed at these frequencies.\\
  \indent Attempts to obtain speeds from the proper motions of jet and counter-jet components indicate
  that these observations are undersampled. The closest pair of images gives apparent speeds of
  0.25 to 0.40c for the jet and 0.17c for the counter-jet. These speeds should be treated as lower
  limits because of possible errors in associating components between epochs. If they are real, they
  indicate that the jet is oriented 30-45\arcdeg~from the line-of-sight and that the component speeds
  along the jet are 0.3-0.5c. Using the jet orientation derived from proper motions, the spectral
  index of the the counter-jet, and a jet-to-counter-jet brightness ratio of 14.4, the inferred bulk
  flow is 0.6-0.7c, which, given the considerable uncertainties in how to measure the brightness ratio,
  is not significantly larger than the component speed.
\end{abstract}


\section{INTRODUCTION}
With the capabilities of Very Long Baseline Interferometry (VLBI), studying jet acceleration and
collimation on sub-parsec scales is achievable, allowing jet collimation theories and general relativistic
magnetohydrodynamic (GRMHD) simulations to be tested. GRMHD simulations have been able to explain the
formation of AGN jets by magnetic fields threading an accretion disk around a super-massive black hole
(SMBH). These fields thread out from the disk in a rotating helical coil, accelerating, guiding, and
collimating the plasma away from the disk. The resulting jet structure is expected to show a large opening
angle near the base and tighter collimation on larger scales \citep[][and references therein]{meier01}.
An example where the large opening angle at the base of the jet may have been observed is in the giant
elliptical galaxy, M87 \citep[Virgo A, 3C\,274, NGC\,4486;][]{junor99}. This observation indicates that
the jet must form within 30 Schwarzschild radii ($R_{\rm s}=2GM_{\bullet}/c^2$) of the black hole (BH)
with an opening angle of at least 60\arcdeg. M87's close proximity, large BH mass
\citep[2-4$\times$10$^9$\,M$_{\sun}$,][]{harms94,macchetto97,marconi97}, and bright jet make it arguably
the best observable candidate to study the jet collimation region. Throughout this paper,
a BH mass of $3\times$10$^9$\,M$_{\sun}$ at a distance of 16\,Mpc \citep{whitmore95,tonry01} is adopted
for M87. A proper motion of 1\,mas yr$^{-1}$ corresponds to an apparent speed of 0.255c, and 0.1 mas
corresponds to a physical distance of 0.0078~pc or 30$R_{\rm s}$.\\
\indent On larger scales (0.1\,pc to 80\,kpc), the jet and radio halo have been intensively observed in the
radio between 90\,cm and 7\,mm with the VLA \citep{biretta95,owen99,owen00} and with VLBI
\citep{reid82,reid89,junor95,biretta02,ly04,dodson06}. It has also been observed in the ultra-violet and
optical with HST \citep{biretta99,perlman03,waters05}, and with \textit{Chandra}
\citep{marshall02,harris03,forman06}.\\
\indent In \S~2, all five 43\,GHz observations of M87 are presented with a description of the data
reduction and imaging process. A description of the structural variations seen between epochs, attempts to
measure proper motions in the jet, the probable detection of the counter-jet, and a discussion of the
opening angle at the jet base are found in \S~3. Finally, \S~4 will summarize the results with remarks on
future high frequency Very Long Baseline Array \citep[VLBA;][]{napier94} studies.

\section{OBSERVATIONS, CALIBRATION, AND IMAGING}
\label{2}
\indent M87 has been observed between 1999 and 2004 with VLBI at 43\,GHz for a total of five epochs,
as summarized in Table~\ref{table1}.\footnotemark[4]
\footnotetext[4]{It has also been observed in 1995, but this observation is excluded from this study
  because of the poor sensitivity to the jet. This observation can be found in \citet{biretta02}.}
An image from epoch 1999.17 has been published in \citet{junor99} and \citet{biretta02}. An image
from epoch 2000.27 has been published by \citet{dodson06}. For epoch 2001.78, M87 was used as a
phase-referencing calibrator to image M84, and an M87 image has been published by \citet{ly04}. 
For epoch 2002.42, M87 was used as a phase-referencing calibrator at multiple frequencies; only the data
at 43 and 22\,GHz will be used here. For epoch 2004.25, M87 was used as a delay calibrator for M84.
All observations used a total bandwidth of 64 MHz with 2-bit sampling.
The first epoch of data was observed with only a single polarization, while the other data were
observed and correlated with full Stokes parameters. The last epoch was the only one for which a
polarization calibration has been attempted. Polarized flux density is detected near the strongest
feature, but due to the short on-source integration time and calibration issues, the results are
not considered reliable. A follow-up movie project, currently in progress, is providing much better
polarization information, which will be discussed when those results are published (Walker, R. C.,
et al., in preparation).

\clearpage
\thispagestyle{empty}
\begin{figure}[htp]
\vspace*{-20mm}
\epsscale{0.5}
\plotone{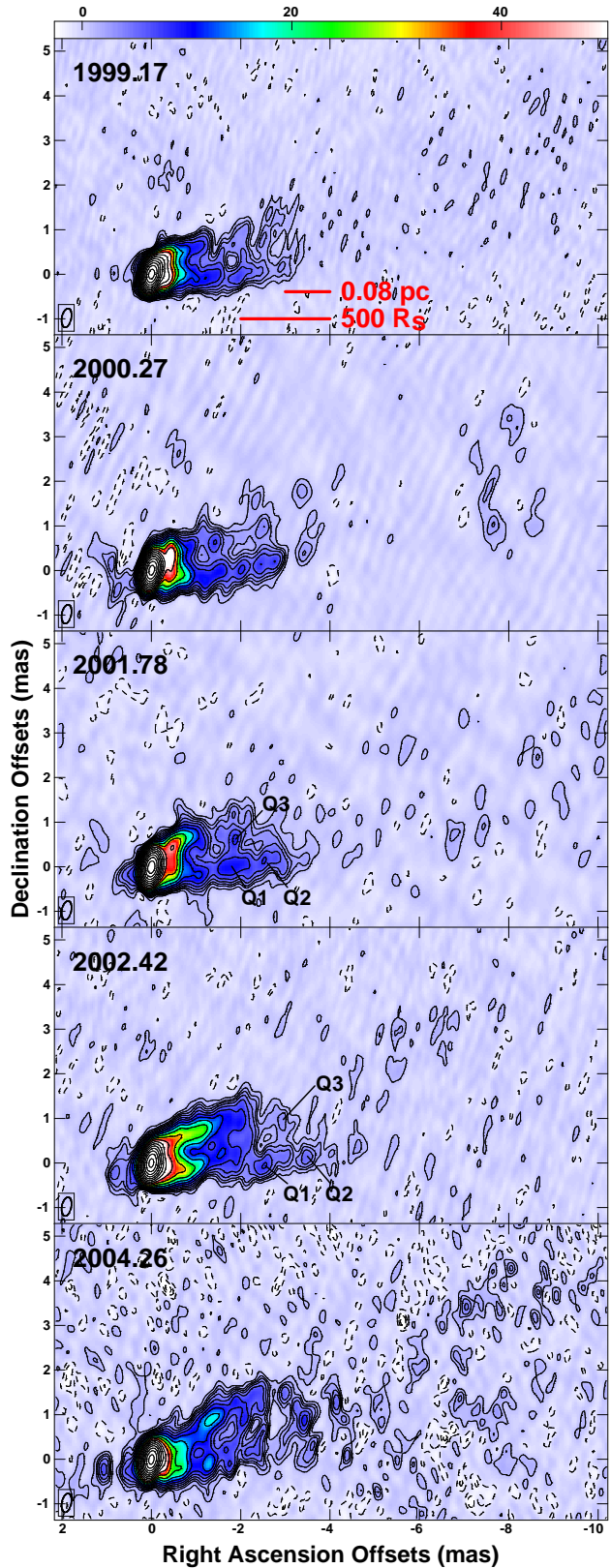}
\caption{Five VLBI 43\,GHz observations of M87 between 1999.17 and 2004.26 with a resolution of
  0.43\,mas~$\times$~0.21\,mas and a PA of $-$11.2\arcdeg. The grey scale (colored in the electronic
  edition) ranges from $-$2.5 to 50\,mJy beam$^{-1}$, and contour levels are $-$2.8, $-$2, $-$1, 1, 2,
  2.8, 4, 5.7, 8, and multiples of $\sqrt{2}$ thereafter until 512 mJy beam$^{-1}$. Jet components
  labeled Q1, Q2, and Q3 in the 2001 and 2002 epochs are those used to obtain proper motions in
  \S\S~\ref{3.3}. A size scale is shown by the thick black lines (red in the electronic edition).}
\label{fig1}
\end{figure}
\clearpage
\begin{deluxetable}{lclrc}
  \tablewidth{0pc}
  \tablecaption{Summary of VLBI Observations}
  \tablehead{
  \colhead{Date}&\colhead{Freq.}&\colhead{Telescopes\tablenotemark{a}}    &\colhead{Int. time}&\colhead{RMS}\\
                &\colhead{GHz}  &                                         &\colhead{min.}     &\colhead{mJy beam$^{-1}$}\\
  \colhead{(1)} &\colhead{(2)}  &\colhead{(3)}                            &\colhead{(4)}      &\colhead{(5)}}
  \startdata
  1999.17       & 43.20         & VLBA+Ef+Mc+Mh+On+Yb+13Y\tablenotemark{c}& 205\tablenotemark{b} & 0.51 \\
  2000.27       & 43.21         & VLBA-Hn                                 &  98\phm{b}        & 0.51 \\
  2001.78       & 43.12         & VLBA                                    &  68\phm{b}        & 0.57 \\
  2002.42       & 22.22         & VLBA-Sc                                 &  23\phm{b}        & 0.64 \\
                & 43.20         & VLBA                                    &  42\phm{b}        & 0.81, 0.51\tablenotemark{d}\\
  2004.26       & 43.12         & VLBA+Ef+Gb\tablenotemark{c}             &  23\phm{b}        & 0.84
  \enddata
  \label{table1}
  \tablenotetext{a}{Instrument abbreviations: Ef=Effelsberg (Germany), Mc=Medicina (Italy),
      Mh=Metasahovi (Finland), On=Onsala (Sweden), Yb=Yebes (Spain), Gb=Green Bank Telescope, and
      13Y=13 VLA antennas.}
  \tablenotetext{b}{Average time for U.S. baselines. Significantly less on global baselines.}
  \tablenotetext{c}{The final image for 1999.17 was processed without Mc and Yb. The final image for 2004.26 was
    processed without Ef and Gb as weather problems degraded the observation.}
  \tablenotetext{d}{The higher RMS is for a lower resolution map with matched resolution to the 22\,GHz image while the
      lower value is for the image that has the same resolution as the other 43\,GHz observations.}
    \tablecomments{A summary of VLBI observations. Col. (1) and (2) list the dates of the observation and
    the observed frequencies in GHz. The instruments used are shown in Col. (3). Col. (4) gives the
    integration time in minutes, and the final column has the off-source RMS in mJy beam$^{-1}$.}
\end{deluxetable}

\clearpage
\indent All observations were reduced using the standard data reduction methods in AIPS. The absolute
amplitude calibrations are those of the original observers and have not all been done in a consistent
manner. For all data sets except the first, opacity corrections were made, the gain normalizations
in self-calibration were based on only those antennas with good conditions, and a lower elevation
limit was set for data used in the gain normalization. With these procedures, the amplitude
calibration should be accurate to better than about 10\%. There appears to have been some difficulty
with the 2000.27 opacity calibration, so that epoch should be treated as having somewhat larger
amplitude errors. The 1999.17 calibration did not include the above steps, and antennas with
{\it a priori} calibration that is typically significantly worse than that of VLBA antennas were
included. The resulting image had a much lower flux density than the other epochs. Examination of the
raw data on selected VLBA baselines indicates that this difference was not justified, so the image was
scaled by a factor of 2.15 to match the average of the other epochs. As far as absolute amplitude is
concerned, 1999.17 should be treated as uncalibrated. Note that none of these calibration offsets affect
the dynamic range or appearance of the images, all of which were re-made from the calibrated data.\\
\indent The images were made in AIPS using the standard procedure involving multiple iterations of
self-calibration and deconvolution. All of the 43\,GHz epochs have been imaged with a resolution of
0.43\,mas~$\times$~0.21\,mas with a PA of $-$11.2\arcdeg. These images are shown in Figure~\ref{fig1}. The
epoch 2000 image differs from what is presented in \citet{dodson06} as a greater effort to image the jet
has been attempted here. The off-source RMS fluctuations for these images are also given in
Table~\ref{table1}.

\section{RESULTS and DISCUSSION}
\label{3}

\subsection{Structural Changes in the Jet}
\label{3.1}
In all five epochs, the edge-brightened jet structure is apparent. One possible interpretation for such a
jet morphology is that the radio jet is associated with a `sheath' of magnetic fields surrounding a much
narrower jet. This has been seen in GRMHD simulations by \citet{koide98} and \citet{nishikawa05} where
two regions, a wide magnetically-driven wind and a narrow gas-driven jet are seen with the former
surrounding the latter. A more recent study by \citet{devilliers05} found two different regions as
well, but with the poloidal magnetic fields in the inner hot, fast outflow rather than the colder,
slower outflow that surrounds the inner outflow. The large scale systematic fields developed despite
initial conditions that included only local random fields in the disk.
The optical and x-ray jets seen on larger scales are much closer to the jet axis than the radio jet
\citep{sparks96,biretta99}. Therefore the optical and x-ray jets may be related to the narrow jet seen in
these numerical simulations.\\
\indent Between the first three epochs, 1999.17, 2000.27, and 2001.78, the structure of the jet is similar
between 1 and 4\,mas west of the core. These epochs show more emission for the southern-half of the
jet while the northern-half of the jet is detected within 2\,mas from the core. There appear to be
structural changes occurring close to the core at about $-$0.5\,mas. The two most recent epochs, 2002.42 and
2004.26, show dramatic changes. Comparatively, the 2002 epoch is more edge-brightened than previous epochs,
as the northern-half of the jet in this epoch remains bright farther from the core than in previous epochs.
Moreover, the 2004 epoch reveals a lower surface brightness for the jet. The jet components that represent
the edge-brightened structure in the 2002 epoch appear to have moved outward in 2004 and become detached
from the inner 1\,mas emission of the jet. It is possible that the poor ($u$,$v$) coverage in the last
epoch may explain these significant changes, and so additional observations are required.

\subsection{The Counter-jet}
\label{3.2}
\begin{figure}[htp]
  \epsscale{0.7}
  \plotone{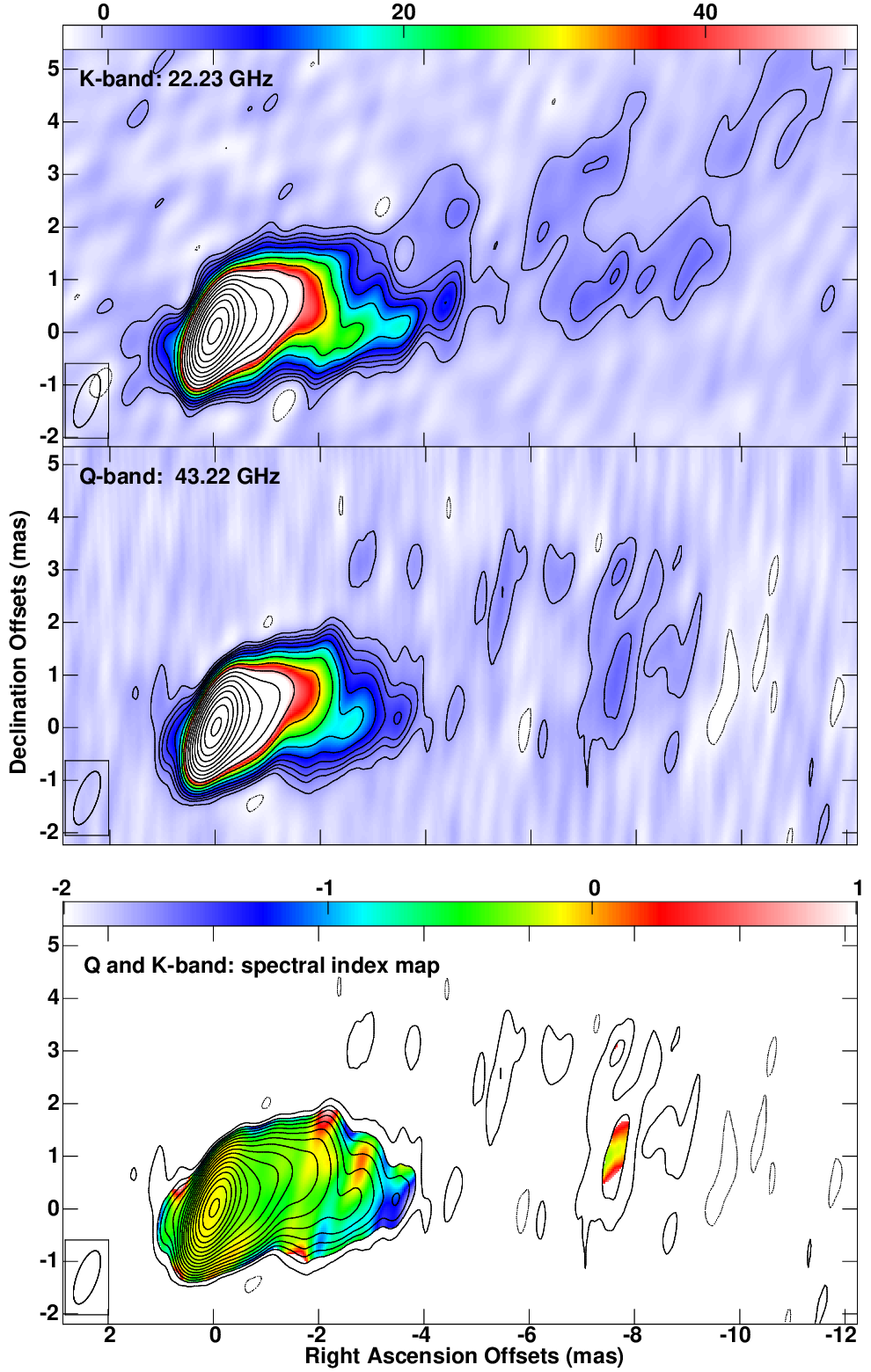}
  \caption{The epoch 2002 22 ({\it top}) and 43\,GHz ({\it middle}) observations. The contour levels are
    $-$2, 2, 4, 5.7, 8, 11.3, 16, and multiples of $\sqrt{2}$ thereafter until 720 mJy beam$^{-1}$, and the
    grey scale (colored in the electronic edition) is from $-$2 to 50 mJy beam$^{-1}$. Both images
    are at a resolution of 1.06\,mas~$\times$~0.38\,mas (PA = $-$20\arcdeg), where a taper was applied to the
    the 43\,GHz observation. The noise in these images are 0.64 (top) and 0.81 mJy beam$^{-1}$ (middle).
    The spectral index map from these images is shown in the bottom panel where the greyscale (colored
    in the electronic edition) is for $\alpha=-2$ to 1, and the 43\,GHz contours are overlaid. The
    spectral index is only shown in locations where the total flux density is above $5\sigma$ in both images.
    The images were aligned by minimizing spectral index gradients across the core, which resulted in a one
    pixel shift (0.035 mas) of on image in both directions.}
  \label{fig2}
\end{figure}

The counter-jet was suggested to exist in the 2001 epoch by \cite{ly04}, and is confirmed by the
two most recent epochs where it appears to move away from the core (see \S\S~\ref{3.3}). It is also
detected in a 22\,GHz observation, which is part of a multi-frequency project for the 2002 epoch. The 22
and 43\,GHz observations for this epoch are shown in Figure~\ref{fig2} where a matched resolution
comparison indicates that both the jet and counter-jet have a spectral index of $\alpha=-0.4\pm0.3$
($S_{\nu} \propto \nu^{\alpha}$). The spectral index map is also included in Figure~\ref{fig2}.
Some counter-jets observed with VLBI are subject to free-free absorption, but that does not appear to
be the case here. Assuming that any free-free absorption has modified the spectral index by less than
0.5 gives a limit on the emission measure of about $7\times10^8$ pc cm$^{-6}$.\\
\indent From an averaged image discussed later (see \S\S~\ref{3.4}), a brightness ratio of 14.4 is found.
There is difficulty in determining a brightness ratio as the width of the counter-jet is much narrower than
the approaching jet. The above brightness ratio was determined by integrating the flux from the jet and
counter-jet at the same distance from the core in a rectangular region. The length of the box along the jet
axis remained the same, but its height varied depending on the width of the jet and counter-jet. Using the
jet orientation derived from proper motions (see \S\S~\ref{3.3}), this brightness ratio suggests that
the bulk flow is 0.6-0.7c, assuming a spectral index of $-$0.4. A much lower brightness ratio
($R=3.4^{+3.1}_{-2.3}$), and hence a lower derived bulk speed, is obtained when comparing just the peak
intensity.\\
\indent To test whether or not the counter-jet is a calibration artifact, several attempts were made
to exclude it during the self-calibration and imaging process for the three 43\,GHz and the single 22\,GHz
observations. After much effort (several self-calibration iterations), the counter-jet was still present,
although reduced. When CLEAN components were again allowed in the counter-jet region, the feature
promptly returned to its previous intensity.\\
\indent Previous radio observations have been unsuccessful at detecting the counter-jet with limits on the
brightness ratio of R=150-1600 \citep[2 cm, 1.2 kpc from the core;][]{biretta89} and $R>200$
\citep[18 cm, 1.2 pc from the core;][]{reid89}. However, some VLBI observations such as the 6\,cm
VSOP+Global Array of \citet{biretta02}, 1.3\,cm observation of \citet{junor95}, and a 21\,cm VLBI
observation by \citet{reid89} appear to have a slight extension of emission east of the core. Also,
emission is seen to the east of the core in one 86\,GHz observation, but not in two other images
\citep{krichbaum06b}. One explanation for why the radio counter-jet is detected only in some observations
is that it is variable partly due to its outward motion. In optical studies, the detection of a bright
hotspot 24\arcsec~away from the core is believed to be caused by a counter-jet interacting with the
interstellar medium \citep{sparks92,stiavelli92}.
\clearpage
\clearpage
\subsection{Proper Motion Measurements}
\label{3.3}
Between the two closest epochs, 2001.78 and 2002.42 (about 7.5 months apart), apparent speeds for jet
components that may be related were found to be between 0.25 and 0.40c. These components are labeled as Q1, Q2,
and Q3 in Figure~\ref{fig1}. Moreover, the counter-jet appears to move eastward from the core at approximately
0.17c. If these speeds are correct, these components would not be present in other epochs. For earlier
epochs, these components would not have originated from the core, and for the last epoch they would have
gone further away and have fallen below the detection limit. This indicates that the year or more
intervals between the other epochs undersample the motions and each image is likely to be of a completely
different set of components. In fact, there is nothing to insure that the components seen in 2001 and 2002
are really the same. For example, VLA and HST observations have found a wide range of speeds from as low
as 0.1c for Knot C \citep{biretta95} to as high as 6c for HST-1 \citep{biretta99}. The highest speeds
would correspond to about 24 mas yr$^{-1}$. Assuming that Q1-3 are related between the two epochs gives
the lowest possible speed and therefore the indicated speeds should be treated as lower limits.\\
\indent Previous VLBI observations have found static or subluminal motions \citep{reid89,junor95,junor00}.
It has been suggested that the presence of subluminal motions for some VLA knots and most VLBI jet
components is likely caused by standing shocks occurring within the jet, and the intrinsic bulk flow of
the jet is represented by the superluminal motions as high as 6c \citep{biretta99}. Evidence supporting
shocks include polarimetric measurements of these knots, and optical and x-ray synchrotron radiation
indicating {\it in situ} particle acceleration to resolve the particle lifetime problem at these higher
frequencies \citep[][and references therein]{perlman05}. Only VLBI observations with shorter time
intervals can determine the speed of jet components unambiguously.\\
\indent With apparent speeds for both the jet and counter-jet, one can estimate the orientation
($\theta$) of the jet axis with respect to the line-of-sight and the intrinsic speed of the components
being observed ($v_{\rm p}$). The brightness is not a factor in this estimate, so the estimate is not
affected by a difference between the bulk speed and the speed of the observed features. The estimate
does depend on an assumption that the jet/counter-jet is a symmetric outflow system. The jet/counter-jet
proper motions, if real, indicate $v_{\rm p}/c$ = 0.3-0.5 and $\theta$ = 30-45\arcdeg. Other studies
have found slightly larger values for $\theta$. For example, \citet{macchetto97} and \citet{marconi97} found
that the circumstellar disk is oriented $\theta$ = 47-65\arcdeg~and $\theta=51$\arcdeg, respectively from
the line-of-sight. The morphologies of VLA radio knots indicate a jet orientation of
$\theta=42.5\pm4.5$\arcdeg~\citep{biretta95}.\\
\indent The bulk flow determined from the brightness ratio in \S\S~\ref{3.2} is somewhat higher than the
component speed.  But alternative ways to measure the brightness ratio give lower values so the difference
should not be considered to be significant. Accurate measurements of the brightness ratio (which is affected
by relativistic beaming) are required to fully constrain the motion and the orientation, but if the jet
orientation and the brightness ratio hold up, this would indicate that the bulk and component speeds are
similar at this frequency. On-going and future observations will provide a better constraint on the
orientation and the speed of the jet.

\subsection{Opening Angle at the Jet Base}
\label{3.4}
\begin{figure*}
\includegraphics[angle=-90, scale=0.70]{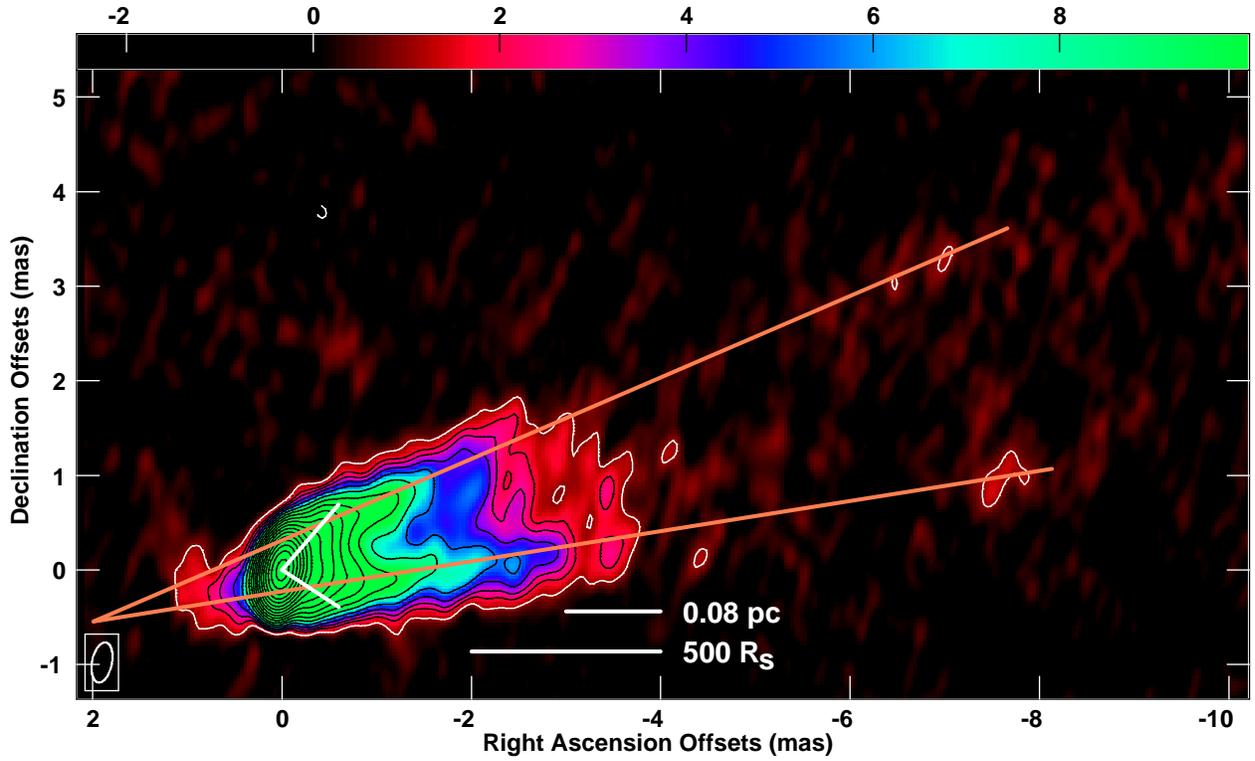}
\caption{An averaged image from all five 43\,GHz observations. The orange lines follow the edge-brightened
  jet emission from 10\,mas west of the core while the white lines represent the wide opening angle
  proposed by \citet{junor99}. The resolution is 0.43\,mas\,$\times$\,0.21\,mas. The color scale is from
  $-$2.5 to 10 mJy beam$^{-1}$  with contour levels of $-$1, 1, 2, 2.8, 4, 5.7, 8, and multiples of $\sqrt{2}$
  thereafter until 512 mJy beam$^{-1}$.}
\label{fig3}
\end{figure*}

An average of all five epochs is shown in Figure~\ref{fig3}. This image shows the edge-brightened structure
out to $\approx$~4\,mas, low-level emission from the jet that is seen out to 10\,mas, and the counter-jet east
of the core. The southern-half of the jet is detected about 1\,mas (0.08\,pc) further than the
northern-half for the brighter inner regions. The two orange lines follow the jet emission where it is
marginally detected out to 10\,mas from the core. These lines intersect at 2\,mas east of the core,
indicating that the opening angle at the base of the jet is wider and collimation occurs further along
the jet, confirming previous conclusions. The white lines are the wide opening angle of
60\arcdeg~reported by \cite{junor99} with higher resolution global VLBI imaging.
\clearpage
\clearpage
There have also been several attempts to image the jet collimation region with mm-VLBI for a higher
resolution image \citep{lobanov00,krichbaum05,krichbaum06a,krichbaum06b}. The former was not successful in
detecting the jet; however, \citet{krichbaum05} indicated that the jet shows a smaller opening angle, which
is in contrast to \citet{junor99}. A small opening angle may be caused by lower sensitivity due to a weaker
jet emission and poor ($u$,$v$) coverage at these frequencies. A more recent 86\,GHz observation of
\citet{krichbaum06b} reveals an edge-brightened jet structure very similar to the five 43\,GHz images
presented in this paper. The southern edge appears to extend further than the northern one, confirming
these 43\,GHz results.\\
\indent Some have offered alternative explanations for an observed large opening angle. One obvious
hypothesis is that the core may not be associated with the brightest point in the jet, but that the core
may be offset by $\sim$2\,mas east (where the orange lines intersect in Figure~\ref{fig3}) of the
bright spot and the jet is invisible until 2\,mas away. This would reduce the opening angle of the jet to
about 15\arcdeg. Also, the counter-jet seen in these observations will be a part of the inner jet rather
than the receding jet with the invisible core interpretation. This introduces a problem as the
eastward motion of the counter-jet is in the wrong direction. However, with the poor time sampling of the
three epochs, it is possible that the counter-jet is not really moving and is part of the inner jet.
\citet{biretta02} argue against this case as in a higher resolution image for the 1999 epoch,
the jet components appear to be directed toward the bright core. Also, if one traces the 22\,GHz jet
emission, one finds that the intersection is 9~mas east from the brightest spot rather than 2~mas at
43\,GHz. However, these arguments favoring the brightest spot as the core are not fully convincing. One
way to possibly resolve this issue is to have a few 43\,GHz phase referencing observations to see if the
`core' moves significantly. However, if the brightest spot does not move, the issue is not resolved
as it could be a standing shock. At lower frequencies, the VLBI core has shown negligible motion of
$0.08\pm0.04$ mas yr$^{-1}$ \citep[][and references therein]{biretta95}.
Perhaps if the counter-jet was more extended in future or lower frequency observations then this
would further support the core being associated with the brightest spot.\\
\indent Another possible explanation is that the jet has a smaller line-of-sight angle closer to the core
than on larger scales. This will cause an underestimation of the true jet size and will reduce the jet
opening angle. \citet{biretta02} argue that this would be improbable given the required geometric
configuration relative to the observer.

\section{CONCLUSIONS}
\label{4}
\indent From these sub-mas resolution observations, the counter-jet for M87 has been confirmed from a
22\,GHz observation and three epochs at 43\,GHz where it appears to move outward. The edge-brightened jet
structure is present in all the 43\,GHz observations. Attempts at proper motion measurements reveal that
lower limits on the apparent speed of the jet are 0.25 to 0.40c and 0.17c for the counter-jet.
Therefore, yearly-sampled epochs cannot provide accurate proper motion measurements. These motions
provide constraints on the jet orientation and the component speeds along the jet to be
30-45\arcdeg~from the line-of-sight and moving at 0.3-0.5c, respectively. The brightness ratio
of 14.4 indicates that the bulk speed is 0.6-0.7c, or somewhat greater than the component speed. But
alternate methods of measuring the brightness ratio give lower values so the difference is not significant.
From both the individual images and an averaged image, a wide
opening angle at the jet base and an edge brightened structure are confirmed.\\
\indent Future work in progress, such as making a properly sampled movie at 43\,GHz for M87 will provide
kinematical information about the jet and counter-jet's components and help localize the core to
determine the opening angle of the jet. In addition, the counter-jet will help determine the
motion of the jet and the angle to the line-of-sight with accurate brightness ratios and proper
motions. Polarization measurements from the extended structure, if possible, will provide a
better understanding of the magnetic fields and help to further constrain jet-collimation models.

\acknowledgements This research has made use of the NASA Astrophysics Data System. W. Junor was
partially supported through NSF AST-9803057. C. Ly acknowledges support through the National Radio
Astronomy Observatory Graduate Research Program. The National Radio Astronomy Observatory is a facility of
the National Science Foundation, operated under cooperative agreement by Associated Universities, Inc. We
thank J. Ulvestad and J. Anderson for providing their uncalibrated data, and an anonymous referee for their
comments that improved the paper. We also thank J. Wrobel for her suggestions that improved this paper
and for her assistance during the planning and observing stages of the 2004 epoch of M84 and M87.

\end{document}